\documentclass[fleqn,usenatbib]{mnras}

\usepackage{mathptmx}

\usepackage[T1]{fontenc}

\DeclareRobustCommand{\VAN}[3]{#2}
\let\VANthebibliography\thebibliography
\def\thebibliography{\DeclareRobustCommand{\VAN}[3]{##3}\VANthebibliography}

\DeclareMathAlphabet{\mathcal}{OMS}{cmsy}{m}{n}

\usepackage{graphicx}	
\usepackage{amsmath}	
\usepackage{amssymb}	
\usepackage{newtxtext,newtxmath}

\usepackage{lipsum}

\usepackage{upgreek}

\newcommand{\msun}{~\mathrm{M}_{\odot}}

\newcommand\iona[2]{#1\,{\sc #2}}

\newcommand{\bs}[1]{\boldsymbol{#1}}
\newcommand{\mat}{\ensuremath{\mathbfss}}

\title[The tails of jellyfish galaxies]{The magnetised and thermally unstable tails of jellyfish galaxies}

\author[Sparre et al. ]{Martin Sparre$^{1,2}$\thanks{E-mail: sparre@uni-potsdam.de}, Christoph Pfrommer$^2$, Ewald Puchwein$^2$
\\
$^{1}$Institut f\"ur Physik und Astronomie, Universit\"at Potsdam, Karl-Liebknecht-Str.\,24/25, 14476 Golm, Germany\\
$^{2}$Leibniz-Institut f\"ur Astrophysik Potsdam (AIP), An der Sternwarte 16, 14482 Potsdam, Germany\\
}

\pubyear{2023}

\begin{document}
\label{firstpage}
\pagerange{\pageref{firstpage}--\pageref{lastpage}}
\maketitle

\begin{abstract}
Jellyfish galaxies are promising \emph{laboratories} for studying radiative cooling and magnetic fields in multiphase gas flows. Their long, dense tails are observed to be magnetised, and they extend up to 100~kpc into the intracluster medium (ICM), suggesting that their gas is thermally unstable so that the cold gas mass grows with time rather than being fully dissolved in the hot wind as a result of hydrodynamical interface instabilities. In this paper we use the \textsc{arepo} code to perform magnetohydrodynamical windtunnel simulations of a jellyfish galaxy experiencing ram-pressure stripping by interacting with an ICM wind. The ICM density, temperature and velocity that the galaxy encounters are time-dependent and comparable to what a real jellyfish galaxy experiences while orbiting the ICM. In simulations with a turbulent magnetised wind we reproduce observations, which show that the magnetic field is aligned with the jellyfish tails. During the galaxy infall into the cluster with a near edge-on geometry, the gas flow in the tail is fountain-like, implying preferential stripping of gas where the rotational velocity vectors add up with the ram pressure while fall-back occurs in the opposite case. Hence, the tail velocity shows a memory of the rotation pattern of the disc. At the time of the nearest cluster passage, ram-pressure stripping is so strong that the fountain flow is destroyed and instead the tail is dominated by removal of gas. We show that gas in the tail is very fragmentative, which is a prediction of \emph{shattering} due to radiative cooling.
\end{abstract}



\section{Introduction}

Galaxy halos are \emph{multiphase} with cold, warm and diffuse gas phases \citep{2017ARA&A..55..389T}. This is, for example, seen in absorption spectroscopy of the circumgalactic medium (CGM) of galaxy samples, where \iona{H}{i} and \iona{O}{vi} ions, corresponding to cold and warm gas, are detected \citep{2014ApJ...792....8W}. A similar diverse set of ions tracing different temperatures are observed in spectroscopy probing gas in the Local Group \citep{2017A&A...607A..48R}.

Realistically treating multiphase gas is a major challenge to modern galaxy formation models relying on hydrodynamical cosmological simulations. To reliably simulate the thermal instability caused by radiative cooling, it is necessary to resolve the \emph{cooling length}, as shown in two- \citep{2018MNRAS.473.5407M} and three-dimensional simulations \citep{2019MNRAS.482.5401S} of a cold-dense cloud interacting with a hot-diffuse wind. For cold CGM gas with a temperature of $10^4$--$10^5$ K it is, however, not possible to resolve this length-scale in state-of-the-art cosmological simulations. This causes gas in the CGM to fragment to the grid-scale of the simulations \citep{2019ApJ...882..156H, 2019MNRAS.482L..85V}. \citealt{2018MNRAS.473.5407M} introduced the term \emph{shattering} to describe the fragmentative nature of gaseous cold clouds with sizes larger than their cooling length that interact with a wind.

Another effect caused by radiative cooling is that large clouds exposed to a hot wind grow, when gas from the surroundings of a cloud cools faster in comparison to the hydrodynamical destruction time-scale of the cloud \citep{2011PASJ...63.1165Y,2016MNRAS.462.4157A,2018MNRAS.480L.111G,2020MNRAS.492.1841L,2020MNRAS.499.4261S,2021MNRAS.501.1143K,2022ApJ...925..199A,2023arXiv230703228A}. This provides a plausible explanation for how cold gas can survive far out in the CGM \citep{2022ApJ...924...82F}.

A promising probe of the physics of multiphase gas is ram-pressure stripping of galaxies infalling into galaxy clusters. In the galaxies' rest frame, the ICM wind interacts with spiral galaxies and provides the ram-pressure to strip gas from the interstellar medium (ISM) to potentially transform the spirals into quenched ellipticals \citep{2022A&ARv..30....3B,2023MNRAS.525.5359M}. As a result of the interaction a long gaseous tail, often detectable in neutral hydrogen \citep[\iona{H}{i};][]{2019MNRAS.487.4580R,2020A&A...640A..22R}, UV \citep{2018MNRAS.479.4126G}, H$\alpha$ \citep{2020ApJ...899...13G}, X-ray \citep{2021ApJ...911..144C} and even molecular gas \citep{2018MNRAS.480.2508M}, is formed. A popular term for such galaxies is \emph{jellyfish galaxies}. Furthermore, \citet{2021ApJ...922L...6F} measured the metallicity of jellyfish tails, as revealed by optical emission lines, and showed that they consist of a mix of stripped ISM from the galaxy and gas from the ICM. Due to their enormous spatial extends of 50--100 kpc \citep{2019ApJ...870...63C,2021ApJ...911..144C} jellyfish tails probe gas shattering and growth of the cold gaseous phase. In addition to probing the effects of radiative cooling and hydrodynamical instabilities, jellyfish galaxies also probe the magnetic field and cosmic ray electrons as shown by \citet{2021NatAs...5..159M}. The authors showed that the orientation of the magnetic field is aligned with the jellyfish tail, which can be understood as the result of \emph{draping} magnetised plasma from a hot wind around the galaxy-wind interface, where the plasma undergoes thermal instability, accreting onto the outer layers of the tail in order to be adiabatically compressed and sheared \citep{2020MNRAS.499.4261S}.

Idealised simulations of the ICM interacting with a galaxy have been used to study jellyfish galaxies from a theoretical point of view. \citet{2016A&A...591A..51S} initialised a gas profile around a cluster and simulated a galaxy orbiting it. An alternative approach to this setup is a \emph{windtunnel} simulation of a galaxy flying through a cluster with a fixed velocity, density and temperature representing a snapshot of ICM conditions. \citet{2009ApJ...694..789T} used such a setup to study the gas loss of jellyfish galaxies in hydrodynamical simulations. The setup was extended to also include a galaxy magnetic field \citep{2014ApJ...795..148T}, which only slightly changes the gas stripping rate of the galaxy, but more importantly leaves a $\upmu$G-strength magnetised jellyfish tail extending far into the ICM. 

In reality, the ICM is, of course, much more complex than a wind with a fixed velocity, density and temperature. Using a time-variable wind velocity and a uniform magnetic ICM wind, \citet{2014ApJ...784...75R} showed that the configuration of the ICM magnetic field influences the tail, and a magnetic wind generally supports the formation a dense tail. \citet{2019ApJ...874..161T}, furthermore, demonstrated that the time-dependent behaviour of the model ICM plays an important role for the evolution of the jellyfish galaxy.

A more realistic geometry of the interaction between a jellyfish galaxy and the ICM can be obtained using cosmological galaxy formation simulations \citep[as in][]{2018ApJ...865..156J,2019MNRAS.483.1042Y,2023arXiv230409202Z,2023arXiv230409199G,2023arXiv230409196R}. A limitation of cosmological simulations in comparison to idealised counterparts is the lower spatial resolution, and their high computational cost. Idealised simulations are also easier to interpret and it is more straightforward to interpret the role of physics model variations.

In this paper we study how the ICM magnetic field influences the jellyfish galaxy tails in idealised simulations. We perform windtunnel simulations with a time-dependent velocity, density, temperature and turbulent magnetic field, all mimicking what a jellyfish galaxy would experience while orbiting a galaxy cluster. The simulations presented in this paper improve on the simulation modelling in \citet{2020MNRAS.499.4261S} and \citet{2021NatAs...5..159M} by including a realistic self-gravitating disc galaxy instead of a spherical non-gravitational cloud. As such, these new simulations enable us to more realistically compare our results to the magnetic field detections of \citet{2021NatAs...5..159M}. We introduce our simulations in Sect.~\ref{Sec:SimulationOverview}, the results are presented in Sect.~\ref{Sec:Results}, in Sect.~\ref{Sec:Discussion} we discuss our results and Sect.~\ref{Sec:Conclusion} summarises our conclusions.

\section{Overview of simulations}\label{Sec:SimulationOverview}

In this subsection, we describe our simulations of a galaxy in a windtunnel with a time-dependent wind model, mimicking a jellyfish galaxy interacting with the ICM.

\subsection{Simulation code and setup}

For our simulations we use the magnetohydrodynamical code \textsc{arepo} \citep{2010MNRAS.401..791S,2016MNRAS.455.1134P,2020ApJS..248...32W}. It uses a Voronoi mesh, which is moving with the gas flow, as this is well suited for astrophysical applications \citep{2012MNRAS.427.2224T,2012MNRAS.425.3024V,2012MNRAS.424.2999S}.

Each simulation has a baryonic target mass, $m_\text{bar}$. If a gas cell is more (less) massive than $2m_\text{bar}$ ($m_\text{bar}/2$) it is refined (de-refined). On top of this we use a \emph{neighbour volume refinement scheme}, which refines a gas cell if it has a volume larger than 8 times that of one of its neighbours. This improves the treatment of regions with large density gradients.

We use a rectangular simulation domain with size $(L_x,L_y, L_z) = (600,1200,600)$ kpc. In the initial conditions (ICs), the galaxy itself is centred on $(x_\text{galaxy,IC},y_\text{galaxy,IC}, z_\text{galaxy,IC}) = (300,450,300)$ kpc. The galaxy is initialised with zero bulk velocity, and the wind moves in the $\hat{\mathbfit{y}}$-direction.

In the lower part of the simulation box we have the \emph{injection region} ($y<46$ kpc), where the ICM wind is created. Here the density, temperature, metallicity and magnetic field components are set to the injection values (as outlined in Sect.~\ref{Sec:ICMWind}).

\begin{table}
\centering
\caption{The table shows the initial conditions for the disc galaxy of mass $M_{200}= 2\times 10^{12}$ M$_\odot$, which will evolve into a jellyfish galaxy in our simulations. Those initial conditions are created with the code {\sc makenewdisk}. We use the same nomenclature as in \citet{2005MNRAS.361..776S}.}
\label{Table:JellyfishGalaxy}
\begin{tabular}{lcl}
\hline\hline
Symbol & Value & Note  \\
\hline
$V_{200}$ & $179.9$ km s$^{-1}$ & Calculated as $\left(10 G H_0  M_{200}\right)^{1/3}$\\
$c$ & $9$ & Halo concentration, $R_{200}/r_{-2}$\\ 
$\lambda$ & $0.0333$ & Halo spin parameter\\
$m_\text{d}$ & $0.041$ & Mass of disc (gas and stars) in units of $M_{200}$\\
$f_\text{gas}$ & 0.1 & Gas fraction of disc, $M_\text{gas}/(M_\text{gas}+M_\star)$\\
$z_0/h$ & $0.2$ & Disc scale height in units of disc scale length\\
$m_\text{HG}$ & 0.1170 & Mass fraction of the gaseous halo\\
\hline\hline
\end{tabular}
\end{table}

\subsection{The ICM wind}\label{Sec:ICMWind}

\subsubsection{Thermal wind properties}

The time-dependent ICM wind, which we input in our simulations, is based on halo A of \citet{2016A&A...591A..51S}. They generated this profile by calculating the orbit of a test particle within a $1.1\times 10^{15}$ M$_\odot$ halo. A gas and halo density profile was assumed for the cluster (see their figure 1), and the initial temperature profile was calculated by assuming hydrostatic equilibrium. A hydrodynamical simulation of a jellyfish galaxy and the ICM was then run, and the density, temperature, and velocity experienced by the jellyfish galaxy was then recorded in time. We note, that the ICM evolved slightly in time.

\begin{figure}
\centering
\includegraphics[width=0.85\linewidth]{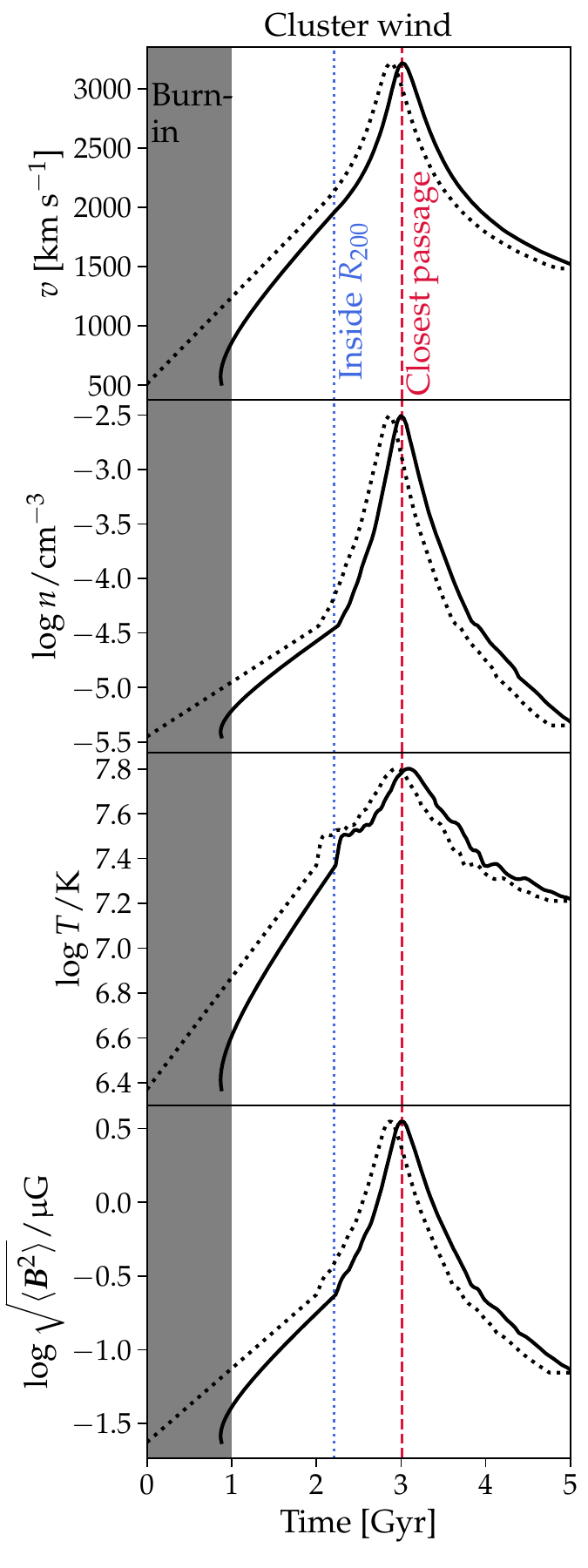}
\caption{In our simulations the jellyfish galaxy experiences a time-dependent wind, which mimics the changing ICM properties along its orbit inside the galaxy cluster. The black-dotted line shows the wind in the lower part of the simulation box, where the wind is injected, and the black-solid line shows the wind at the location of the galaxy taking into account the time it takes for the wind to move from the lower part of the box to the galaxy centre. In the first Gyr, the wind has not yet reached the galaxy, so the galaxy can settle into an equilibrium before ram-pressure stripping becomes important. We refer to this as the \emph{burn-in} phase. At a time of $2.2$ Gyr, the galaxy enters $R_{200}$ and we switch to the wind profile of our galaxy cluster model (from \citealt{2016A&A...591A..51S}), with the nearest passage of the cluster core at $3.0$ Gyr. The amplitude of the magnetic field is chosen such that we obtain $\beta \equiv P_\text{th}/P_{B}=50$ for the wind.}
\label{Fig10_PlotClusterWind}
\end{figure}

We base the wind profiles of our simulations on these resulting profiles. In the first 2 Gyr, we have extrapolated their profiles to allow a \emph{burn-in phase}, where our galaxy settles into equilibrium before it is hit by the wind. We define our burn-in phase to be the first 1 Gyr of the simulation. Our resulting velocity, density and temperature profiles are shown in Fig.~\ref{Fig10_PlotClusterWind}. We show both the wind-profile, which is read into the \emph{injection region} (dashed line), as well as the wind experienced by the galaxy (solid line). The latter is calculated by adding a time-offset of $y_\text{galaxy, IC}/\varv_y(t)$. Here $y_\text{galaxy, IC}=450$ kpc is the initial $y$-coordinate of the galaxy centre.

In Fig.~\ref{Fig10_PlotClusterWind}, we have marked the time, when the galaxy enters the virial radius of the galaxy cluster, $R_{200}$, which is the radius interior to which the mean density equals 200 times the critical density of the universe, as well as the time of the nearest core passage, i.e., where the density, temperature and velocity profiles peak. In the wind we set the metallicity to Z$_\odot/3$ with a solar abundance pattern \citep{2009ARA&A..47..481A}.

\subsubsection{The turbulent magnetic field}\label{SecBturb}

The time-dependent magnitude of the magnetic field is chosen to ensure an initial magnetic--to--thermal pressure ratio of $\beta = 50$ (see Fig.~\ref{Fig10_PlotClusterWind}). We generate magnetic field fluctuations on a periodic cube with a side length of 600 kpc, and a turbulent power spectrum (generated with the code from \citealt{2018MNRAS.481.2878E}) of the form,
\begin{align}
P(k) =  \left|\tilde{B}_i (k)\right|^2 =
\begin{cases}
A, &\text{ for } k < k_\text{inj},\\
B k^{-11/3} \exp \left( - (k/k_\text{crit})^2 \right), &\text{ for }  k \geq k_\text{inj},
\end{cases}
\end{align}
with $k_\text{inj}=1/50$ kpc$^{-1}$ and $k_\text{crit}= 1/6$ kpc$^{-1}$. The adopted turbulent injection scale\footnote{Note that we use a convention such that $k_x \equiv 1/x$ and similar for the other dimensions.} of $\ell_\rmn{inj}=50$~kpc is a compromise of various inferences from astrophysical observations: deprojecting Faraday rotation measure fluctuations in the centre of the Hydra A cool core cluster yields a lower limit on the injection scale of magnetic turbulence of $\ell_\rmn{inj}>8$~kpc \citep{2011A&A...529A..13K}. There are no measurements of $\ell_\rmn{inj}$ on larger scales so that we have to infer the magnetic injection scale from that of kinetic turbulence. Assuming a fluctuating magnetic dynamo necessarily implies an outer scale of magnetic turbulence that is smaller than that of kinetic motions by a factor that depends on the magnetic Prandtl number \citep{2005PhR...417....1B}; this behaviour is also observed for fluctuating magnetic dynamos during galaxy formation with radiative gas physics \citep{2017MNRAS.469.3185P,2022MNRAS.515.4229P}. Mapping out the line-of-sight velocity and (turbulent) velocity dispersion of the X-ray emitting ICM in galaxy clusters demonstrates the presence of velocity gradients (and possibly eddies) across scales of $\sim80$~kpc \citep{2016Natur.535..117H}. Using a number of assumptions, the kinetic power spectrum can be inferred from thermal pressure fluctuations as observed in the X-rays to yield $\ell_\rmn{inj,kin}\gtrsim90$~kpc in the Coma cluster \citep{2004A&A...426..387S}. Deprojecting X-ray surface brightness fluctuations yield characteristic fluctuation length scales ranging from 10 kpc (in the centre) to 30 kpc (at cluster centric radii of $\sim200$~kpc) \citep{2015MNRAS.450.4184Z} that can be pushed to $\ell_\rmn{inj,kin}\gtrsim300$~kpc at intermediate radii in the Perseus Cluster \citep{2023MNRAS.518.2954D}.

The exponential cutoff scale is chosen so that we well resolve all turbulent fluctuations throughout our simulations. We note that the density of our ICM wind is time-dependent following Fig.~\ref{Fig10_PlotClusterWind}, so the spatial resolution of our wind also varies in time (because we use the same target gas mass throughout the simulation).

%
%


\begin{table*}
\centering
\begin{tabular}{lc|ccl}
\hline\hline
name (1)  & $m_\text{bar}/\msun$ (2)  & inclination $i$ (3)  & Wind's magnetic field (4) & Note (5) \\
\hline
{\tt 90deg-Bturb }  & $3.16\times 10^5$ & 90$^{\circ}$    &  Turbulent &Cluster wind hits galaxy \emph{edge-on} \\
{\tt 60deg-Bturb }   & $3.16\times 10^5$  & 60$^{\circ}$ &  Turbulent &  \\
{\tt 60deg-NoB }   & $3.16\times 10^5$  & 60$^{\circ}$ &  0 &  non-MHD simulation of {\tt 60deg-Bturb} \\
{\tt 60deg-Bx }   & $3.16\times 10^5$  & 60$^{\circ}$  &  $\bs{B}_\text{wind}$ is along $\hat{\bs{x}}$ & $\bs{B}_\text{wind}$ is perpendicular to $\bs{\varv}_\text{wind}$   \\
{\tt 60deg-Bturb-HR}   & $3.95\times 10^4$  & 60$^{\circ}$  &  Turbulent   & A high-resolution version of {\tt 60deg-Bturb}  \\
{\tt 30deg-Bturb }  &  $3.16\times 10^5$  & 30$^{\circ}$  &  Turbulent & \\
{\tt \phantom{0}0deg-Bturb}  & $3.16\times 10^5$  & 0$^{\circ}$  &  Turbulent & Cluster wind hits galaxy \emph{face-on}\\
\hline\hline
\end{tabular}
\caption{An overview of simulation parameters. In all simulations the wind velocity is in the $\bs{y}$-direction. We show the simulation name (1), the baryonic mass resolution (2), the galaxy disc's inclination $i\equiv |\measuredangle (\bs{L}_\text{disc},\bs{\varv}_\text{wind})|$ (3), the ICM wind's magnetic field configuration (4) and an explanatory note (5).}
\label{Table:Simulations}
\end{table*}

\subsection{Initial conditions for the galaxy}\label{GalaxyICs}

We use the software {\sc makenewdisk} \citep{2005MNRAS.361..776S} to create the ICs for a disc galaxy with a halo mass of $M_{200}= 2\times 10^{12}$ M$_\odot$. This is the galaxy experiencing ram-pressure stripping by the ICM wind. The input parameters are summarised in Table~\ref{Table:JellyfishGalaxy}. We calculate the virial velocity $V_{200}$, which is an input parameter of {\sc makenewdisk}, based on the above choice of $M_{200}$. We use the approximation $V_{200}\simeq \left(10 G H_0  M_{200}\right)^{1/3}$ and use $H_0 = 0.6774$ km s$^{-1}$ Mpc$^{-1}$ from the Planck15 cosmology \citep{2016A&A...594A..13P}.

A coordinate system with $x^\prime \equiv x - x_\text{galaxy,IC}$ (and similarly for $y^\prime$ and $z^\prime$) is used to describe the initial properties of the galaxy.

The dark matter profile is described as a Hernquist profile \citep{1990ApJ...356..359H}, mimicking an NFW profile \citep{1996ApJ...462..563N} with a halo concentration of $c=R_{200}/r_{-2}$,
where $r_{-2}$ is the radius where  $\text{d}\log \rho/\text{d}\log r=-2$, i.e. the scale radius of the NFW profile.

{\sc makenewdisk} creates a stellar and gaseous disc with surface density profiles following
\begin{align}
\Sigma_\star &= \frac{M_{\star\text{,disc}}}{2\pi h^2} \exp \left( -\frac{\mathcal{R}}{h}\right),\\
\Sigma_\text{gas} &= \frac{M_{\text{gas,disc}}}{2\pi h^2} \exp \left( -\frac{\mathcal{R}}{h}\right),\label{SigmaGas}
\end{align}
where $\mathcal{R}\equiv \sqrt{x^{\prime 2}+y^{\prime2}}$ is the cylindrical radius and $h$ is the exponential scale length. For the stellar component, this surface density profile is achieved by sampling from the density profile,
\begin{align}
\rho_\star (\mathcal{R},z^\prime) \equiv \frac{M_{\star\text{,disc}}}{4\pi z_0 h^2} \text{sech}^2 \left( \frac{z^\prime}{2 z_0} \right) \exp \left( - \frac{\mathcal{R}}{h}\right),
\end{align}
where $z_0$ is the scale height of the disc. We choose $z_0=0.2h$, and $h$ itself is determined based on the halo spin parameter ($\lambda$) and assuming the disc to be centrifugally supported (solving equation (9) in \citealt{2005MNRAS.361..776S}). For our galaxy, our parameters displayed in Table~\ref{Table:JellyfishGalaxy} yield a value of $h=4.16$ kpc, which is comparable to high-resolution simulations of Milky-Way-mass halos \citep{2017MNRAS.467..179G,2020MNRAS.498.2968L}. 

The effective equation of state, self-gravity and the choice of a surface density profile give no freedom for a $z$-dependence of the disc (see \citealt{2005MNRAS.361..776S} for details). For the gaseous disc, the density profile is determined  by (i) assuming a central density in the $z^\prime=0$ \emph{mid-plane}, and (ii) integrating the density in the $z$-direction to obtain $\Sigma_\text{gas}$. The central density in the first step is adjusted and the process is repeated until the surface density profile in Eq.~\eqref{SigmaGas} is obtained. This process also determines the pressure (and temperature) of the gas disc, because the gas in the disc is star-forming and is described by an effective equation of state (see Sect.~\ref{Sec:GalFormPhysics}) where the effective thermal pressure is a function of density $P_\text{therm}=P_\text{therm}(\rho)$.

The disc has a mass of $M_{\star\text{,disc}} + M_{\text{gas,disc}} = m_\text{d} M_{200}$, and it has a gas fraction of $f_\text{gas}\equiv M_{\text{gas,disc}}/(M_{\star\text{,disc}} + M_{\text{gas,disc}})$. We also assume that the fraction of the total angular momentum in the disc is $m_\text{d}$. We include a gaseous halo with a mass of $m_\text{HG}M_\text{200}$. It follows a density profile of the same shape as the dark matter density profile, and the temperature of the halo gas is calculated assuming hydrostatic equilibrium. We initialise the gaseous halo with a spin parameter identical to the dark matter spin parameter ($\lambda$ in Table~\ref{Table:JellyfishGalaxy}). For the ICs we have chosen a total baryon fraction $(M_\text{HG}+M_\text{d})= 0.158$, which agrees with constraints from the Planck 2015 results (\citealt{2016A&A...594A..13P}; we use the implementation of the Planck15 cosmology in the \citealt{2022ApJ...935..167A} python package). We neither initialise a bulge nor a black hole.

For the galaxy itself, we initialise a magnetic field following the profile,
\begin{align}
B_x &= - B_0 \frac{y^\prime}{r+10^{-12} R_{200}} \exp \left( -\frac{\mathcal{R}}{h} \right)   \exp \left( -\frac{|z^\prime|}{z_0} \right), \label{Beq1}\\
B_y &= B_0 \frac{x^\prime}{r+10^{-12} R_{200}} \exp \left( -\frac{\mathcal{R}}{h} \right)   \exp \left( -\frac{|z^\prime|}{z_0} \right), \label{Beq2}\\
B_z &= B_\text{min}.\label{Beq3} 
\end{align}
Here $B_0=50$ $\upmu$G is the central magnetic field, and $B_\text{min}=2.40 \times 10^{-8}$ $\upmu$G is a small floor value of the magnetic field at large distances. The term $10^{-12} R_{200}$ is added to avoid division by zero in the galaxy centre. This field is divergence free by construction. We initialise the gas in the disc as having solar metallicity and solar abundances. The gas in the halo has a metallicity of $1/3$ solar.

\subsection{The simulations: varying the resolution and the wind's inclination}

Our simulations are summarised in Table~\ref{Table:Simulations}. For our standard resolution level, we use a baryonic mass resolution of $m_\text{bar}=3.16\times 10^5\msun$. At this resolution we perform simulations, where the galaxy experiences the wind at an inclination of $i=90^{\circ}$ (edge on), $60^{\circ}$, $30^{\circ}$ and $0^{\circ}$ (face on). We name these simulations {\tt 90deg-Bturb}, {\tt 60deg-Bturb}, {\tt 30deg-Bturb} and {\tt 0deg-Bturb}, respectively, where the postfix {\tt Bturb} indicates that we use a turbulent magnetic field in the ICM wind as described in Sect.~\ref{SecBturb}. 

When creating the initial conditions for the galaxy (Sect.~\ref{GalaxyICs}) we obtain a disc with the angular momentum pointing in the $\hat{z}$-direction. To change the inclination by 30, 60 or 90 degrees relative to the wind (pointing in the $\hat{y}$-direction), we rotate the galaxy around the $x^\prime$-axis, and hence obtain a disc rotation axis pointing along $\mat{R}_{x^\prime}(30^\circ)\hat{\bs{z}}^\prime=-\frac{1}{2}\hat{\bs{y}}^\prime+\frac{\sqrt{3}}{2}\hat{\bs{z}}^\prime$, $\mat{R}_{x^\prime}(60^\circ)\hat{\bs{z}}^\prime=-\frac{\sqrt{3}}{2}\hat{\bs{y}}^\prime+\frac{1}{2}\hat{\bs{z}}^\prime$ and $\mat{R}_{x}^\prime(90^\circ)\hat{\bs{z}}^\prime=-\hat{\bs{y}}^\prime$, respectively. Here $\mat{R}_{x^\prime}$ is the rotation matrix describing a rotation around the $x^\prime$-axis,
\begin{align}
\mat{R}_{x^\prime}(\theta) = 
\begin{pmatrix}
1 & 0 &  0 \\
0 & \cos \theta &  -\sin \theta  \\
0 & \sin \theta &  \cos \theta 
\end{pmatrix}.
\end{align}
We also appropriately rotate the galaxy magnetic field (described in Eqs.~\ref{Beq1}-\ref{Beq3}).

To test the role of the magnetic field orientation of the ICM wind, we perform a single simulation ({\tt 60deg-Bx}) with a uniform wind magnetic field aligned with the $x$-axis. We also perform a simulation with MHD completely disabled ({\tt 60deg-NoB}); i.e. a pure hydrodynamical simulation. Furthermore, we perform a single simulation at a 8 times higher mass resolution with a target mass of $3.95\times 10^4\msun$, which we regard as our \emph{flagship} simulation.

\begin{figure*}
\centering
\includegraphics[width=0.7\linewidth]{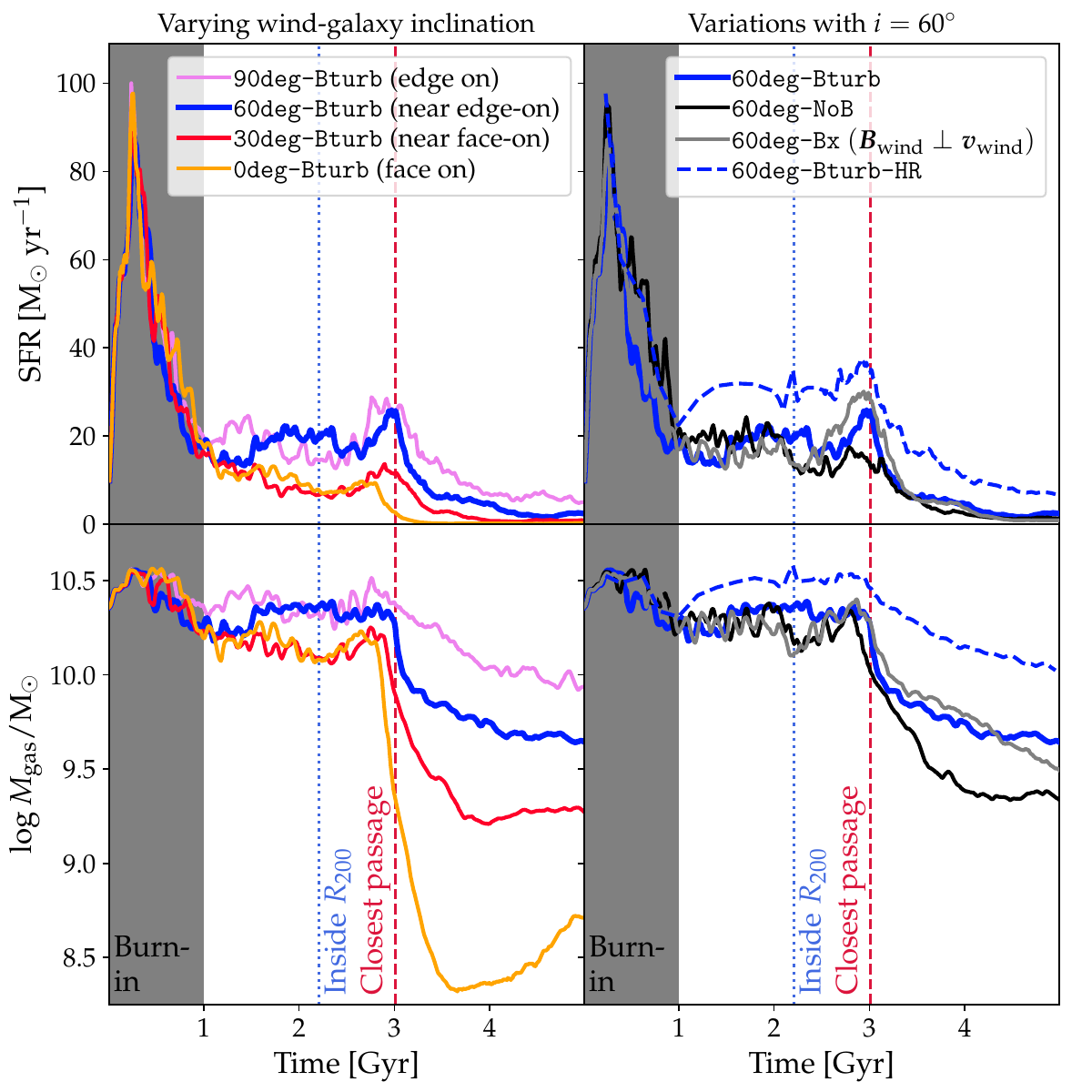}
\caption{Star formation rate (SFR, upper panels) and gas mass (lower panels) measured within twice the stellar half mass radius. \emph{Left panels}: We show the effect of varying the inclination of the galaxy relative to the wind. Gas is more efficiently removed during the simulated cluster core passage as well as for (nearly) \emph{face-on} disc geometries. This is reflected both by the SFR and $M_\text{gas}$ after $3$--$4$ Gyr. \emph{Right panels}: We vary the magnetic field configuration (i.e., we adopt a turbulent vs.\ a homogeneous field vs.\ no magnetic field) and show results from our higher resolution simulation. The magnetic field better protects the gas from stripping and delays the process while the higher-resolution simulation can hold on to its gas for much longer.}
\label{Analyse8_MassEvolutionEvolution}
\end{figure*}

\subsection{Galaxy simulation physics} \label{Sec:GalFormPhysics}

The physical processes in our simulations rely on the galaxy formation model from the Auriga simulations \citep{2017MNRAS.467..179G} with the main differences being that our simulations are not cosmological, and do not include black holes or AGN feedback.

We use ideal MHD and use the implementation described in \citet{2011MNRAS.418.1392P} and \citet{2013MNRAS.432..176P}. We model the ISM using the model of \citet{2003MNRAS.339..289S}, where gas with a density above the star formation threshold density, $n_\text{sf}=0.157$ cm$^{-3}$, is described by an effective equation of state that is derived from a multiphase ISM, which has a contribution from a $10^3$ K cold phase and a hot phase. The physical processes governing these phases are formation of stars (stellar population particles) from the cold phase, \emph{evaporation} of the cold gas by type II supernovae and radiative cooling of the hot phase.

Our modelling of stellar winds is slightly different than in the Auriga simulations, where the wind velocity is calculated based on the local dark matter velocity dispersion. Instead, we set the mass loading factor to a value of $\eta = 2$ and assume energy-driven winds following equation~(6) in \citet{2013MNRAS.428.2966P}. We obtain a wind velocity of 1003 km s$^{-1}$. We note, that a constant value of $\eta$ (and hence also the wind velocity) would cause too strong outflows in low-mass galaxies in a simulation of a representative cosmic volume with a complete population of galaxies with different masses. By contrast, in our paper we focus on a single galaxy and are therefore not affected by such scaling problems.

We follow radiative cooling using the UV background from the model in \citet{2009ApJ...703.1416F} at $z=0$. For UV self-shielding we use prescriptions from \citet{2013MNRAS.430.2427R}. We use a Chabrier IMF \citep{2003PASP..115..763C} for all calculations. An overview of how metals are produced in the galaxy and distributed to the gas can be found in \citet{2013MNRAS.436.3031V}.

\section{Results}\label{Sec:Results}

\subsection{Star formation and gas mass}

We start out showing how the ram-pressure stripping by the ICM wind affects integral properties of the gas in the galaxy. In Fig.~\ref{Analyse8_MassEvolutionEvolution}, we show the star formation rate (SFR) and the gas mass ($M_\text{gas}$). Both quantities are limited to gas within two times the stellar half mass radius.

\subsubsection{Simulations with varying inclination}

We start by describing the simulations at our standard resolution with a turbulent magnetic ICM wind i.e., the simulations {\tt 90deg-Bturb} (edge on), {\tt 60deg-Bturb} (nearly edge-on), {\tt 30deg-Bturb} (nearly face-on) and {\tt 0deg-Bturb} (face on), where we vary the inclination (left panel in the figure). In the initial burn-in phase there is a burst of star formation. Here, the SFR behaves nearly identical in all simulations, implying that ram-pressure stripping is not yet playing any role. Otherwise, we would see an inclination-dependent effect of the stripping. The ICM injection density and velocity is also constructed to be low in the burn-in phase. Instead the burst is caused by the ICs evolving into an equilibrium. We note there are stellar winds associated with this burst, which enriches the gaseous halo. 

After $\gtrsim 1.5$ Gyr the face-on and the $i=30^\circ$ simulations have a slightly lower SFR and gas mass in comparison to the other simulations. This is a signature of ram-pressure stripping, because the cross section area is largest at low inclinations (the cross section area is proportional to $\cos i$). Such an inclination-dependence of SFR and gas mass is also seen in the simulations of \citet{2020ApJ...905...31L} and \citet{2023arXiv230109652A}.

Near (or slightly before) the closest passage of the cluster core, there is an increase in SFR in all simulations. This is consistent with the similar simulations from \citet{2023arXiv230907037Z}. The peak SFR value is largest in the edge-on simulation, because gas is pushed into the centre of the galaxy by the ram pressure. As the inclination decreases (and the galaxies are oriented more face-on) the strength of this burst declines, and the gas mass is also stripped faster.

After the central passage the galaxy gas mass has been lowered by a factor of $\gtrsim 10$ (in comparison to the gas mass when the galaxy was at $R_{200}$) for the face-on and $30^\circ$ simulations, and their SFRs are similarly suppressed. These galaxies are therefore undergoing quenching, whereas the galaxies with larger inclinations maintain larger SFRs throughout the simulations. Whether a galaxy is quenched of course depends on the quenching definition, which is a somewhat arbitrary choice, but our simulations are certainly consistent with ram-pressure stripping being able to cause quenching.

\begin{figure*}
\centering
\includegraphics[width=0.9\linewidth]{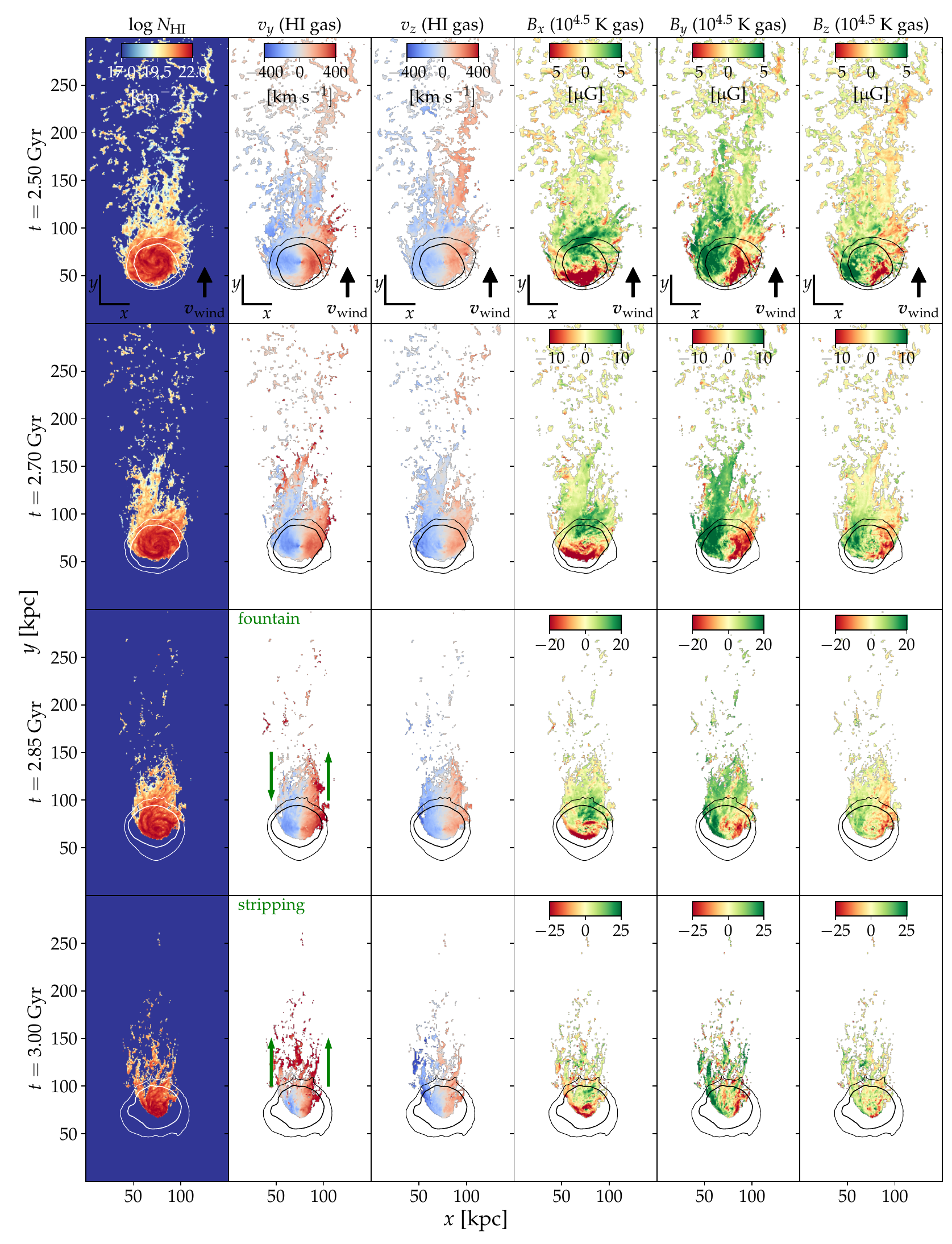}
\caption{Projection maps of the high-resolution simulation, {\tt 60deg-Bturb-HR}. Before the nearest cluster core passage (at $t\leq 2.85$ Gyr) there is a dipole structure in the tail of $\varv_y$ and $\varv_z$. At the nearest passage (at $t=3.0$ Gyr), where the strongest ram-pressure is experienced, stripping is very violent and the central parts of the galaxy are affected. Here $\varv_y$ in the tail no longer has a dipole structure and instead shows clear signatures of ISM stripping. The magnetic field component along the tail ($B_y$) is larger in comparison to the perpendicular directions ($B_x$ and $B_z$).}
\label{Analyse10_Multipanel_vandB_testsim2HR_RotMatrix30deg_080_Movie}
\end{figure*}

\begin{figure}
\centering
\includegraphics[width=\linewidth]{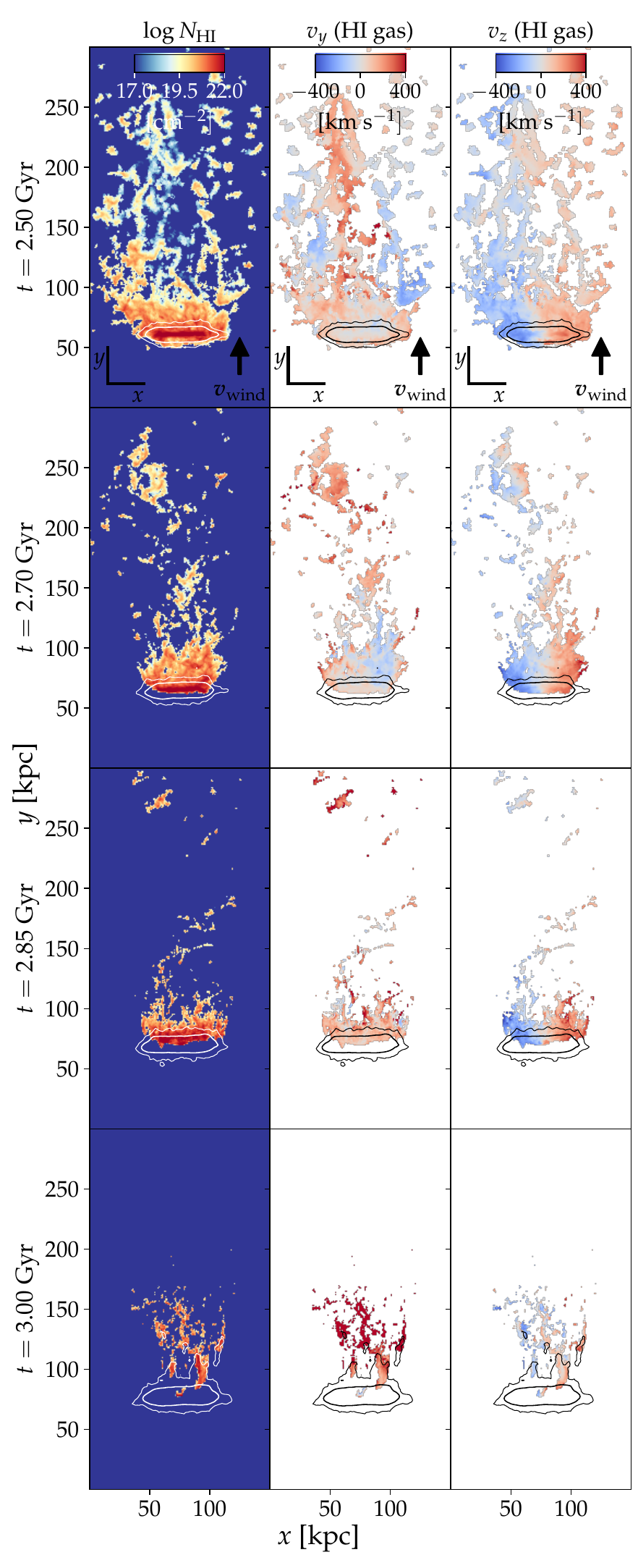}
\caption{\iona{H}{i} column density and velocity components similar to Fig.~\ref{Analyse10_Multipanel_vandB_testsim2HR_RotMatrix30deg_080_Movie}, but for the simulation where the wind hits the galaxy in a face-on projection ({\tt 0deg-Bturb}). For the $\varv_y$-component there is no sign of a fountain flow connected to the rotation flow of the galaxy, as we saw for {\tt 60deg-Bturb-HR} in Fig.~\ref{Analyse10_Multipanel_vandB_testsim2HR_RotMatrix30deg_080_Movie}.}
\label{Analyse10_Multipanel_vandB_FaceOnGeometryNoB_4-FaceOn_300_Movie}
\end{figure}

\subsubsection{The orientation of the magnetic ICM wind}

We now describe the simulations with different magnetic field configurations, shown in the right panel of Fig.~\ref{Analyse8_MassEvolutionEvolution}. All these simulations have an inclination of 60$^\circ$.

The simulation with an initial uniform magnetic field that is perpendicular to the wind velocity shows a SFR and gas mass  very similar to that in the corresponding simulation with a turbulent magnetic wind (simulation {\tt 60deg-Bx} and {\tt 60deg-Bturb}, respectively). The magnetic field configuration does therefore not strongly influence these integral measures of the galaxy, but as we will see in Sect.~\ref{Sec:MagnetigFieldAlignment}, it will change the magnetic structure of the tail of the galaxy.

The simulation without a magnetic field also has a similar evolution as the fiducial simulation ({\tt 60deg-NoB} and {\tt 60deg-Bturb}, respectively). The most visible differences are that the strength of the starburst is slightly weaker for the simulation without MHD at the time of the closest passage, and the final gas mass is also lower without magnetic fields.

\subsubsection{Spatial resolution}

The high resolution simulation shows a higher SFR in comparison to the same simulation with our standard resolution ({\tt 60deg-Bturb-HR} and {\tt 60deg-Bturb}, respectively). This is a well-known result of our subgrid ISM model, which tends to under-estimate the SFR if the gradients within the disc are not well-resolved. This is also seen in the Auriga simulations \citep[][see their fig. 23]{2017MNRAS.467..179G} and in the merger simulations of \citet{2016MNRAS.462.2418S}. We note, that the evolution of the SFR are qualitatively similar for these two simulations, and only the actual normalisation is different.

\subsection{Gas maps}\label{subsec:gasmaps}

In Fig.~\ref{Analyse10_Multipanel_vandB_testsim2HR_RotMatrix30deg_080_Movie}, we show gas maps visualising our high-resolution simulation, {\tt 60deg-Bturb-HR}. We show the projected \iona{H}{i} column density, the \iona{H}{i} weighted mean of the $\varv_y$ and $\varv_z$ velocity, and the magnetic field components. We calculate the \iona{H}{i} content of each gas cell following the tables used in \citet{2018MNRAS.475.1160H}, which are based on CLOUDY \citep{1998PASP..110..761F,2017RMxAA..53..385F} and take into account self-shielding, collisional excitation and a radiation field from the UV background \citep{2009ApJ...703.1416F}. We define star-forming gas cells to have a temperature of $10^4$ K, so they contribute to the neutral phase. For a full description see \citet{2018MNRAS.475.1160H}.

We project the magnetic field along the $z$-axis by calculating,
\begin{align}
\langle B_j\rangle_z \equiv \frac{\int  B_j(z) G\left( \log \frac{T(z)}{\text{K}}\right) \rmn{d}z}{\int G\left(\frac{\log T(z)}{\text{K}}\right)  \rmn{d}z},
\end{align}
where $j=x,y$ or $z$ and for the weighting function, $G$, we use a Gaussian with a mean of $\log (T/\text{K})=4.5$ and a standard deviation of 0.5 dex.  With this definition, we obtain a projected map of the magnetic field that is anchored in gas with a temperature of $10^{4.5\pm 0.5}$ K. This choice is made to select a similar gas reservoir as traced by \iona{H}{i} (we shy away from directly calculating the \iona{H}{i} averaged magnetic field, which cannot be measured observationally, and instead we use the before-mentioned log-normal selection). The line-of-sight-integral is calculated numerically by discretising it into points separated by 0.25 kpc.

\subsubsection{HI column density}

All the snapshots shown have remarkable \iona{H}{i} tails, which identifies this galaxy as a jellyfish galaxy. The tail's \iona{H}{i} column density peaks at a value of $\sim 10^{20}$ cm$^{-2}$, which is also seen in observed jellyfish galaxies \citep{2019MNRAS.487.4580R,2020A&A...640A..22R}. Visually, the \iona{H}{i} tail appears clumpy and fragmented, which could be an indication of \emph{shattering} as a result of thermal instability -- we study this further in Sect.~\ref{subsec:shattering}.

The existence of a dense tail reaching far downstream shows that the clouds constituting the tail are not destroyed by hydrodynamical instabilities.

\subsubsection{Kinematics}

At the times prior to the nearest passage (at $t\leq 2.85$ Gyr), the $\varv_y$ and $\varv_z$ velocity components reveal a dipole distribution, where the rotation pattern of the disc is reflected in the tail. The dipole pattern in $\varv_y$ implies that a part of the stripped gas undergoes a fountain flow, where it is re-accreted back to the disc. Such a \emph{fallback} indicating a fountain flow has been observed for the molecular gas in a ram-pressure-stripped galaxy \citep{2021ApJ...921...22C}. The finding that the tail's gas also has memory of $\varv_z$ is a result of angular momentum conservation.

At the time of the nearest passage in the galaxy cluster, the $\varv_y$ velocity pattern changes, and the fountain flow is no longer visible. Instead, stripping is very efficient. At the same time, gas within the disc is compressed (gas slightly upstream is pushed towards the centre) and at the same time producing a ram-pressure induced starburst (as seen in Fig.~\ref{Analyse8_MassEvolutionEvolution}).

A fountain flow with a $\varv_y$-value corresponding to the rotation pattern of the galaxy is seen in all simulations at some point, except for the situation where the wind hits the galaxy face on. For this simulation we show the \iona{H}{i} column density and kinematical features in Fig.~\ref{Analyse10_Multipanel_vandB_FaceOnGeometryNoB_4-FaceOn_300_Movie}. The rotation of the galaxy gives zero contribution to the $\varv_y$-component, so there is no rotation signal to be extended downstream. On the other hand, the $\varv_z$-component in the tail clearly connects to the rotation pattern of the galaxy.

A necessity for the fountain flow, which we identified in Fig.~\ref{Analyse10_Multipanel_vandB_testsim2HR_RotMatrix30deg_080_Movie}, to exist is a misalignment of the (unsigned) angular momentum vector of the disc with the wind direction.

\subsubsection{Magnetic field}

We expect the magnetic field in the tail to be influenced by the behaviour of ICM wind as well as the gas stripped from the galaxy itself as the downstream tail consists of a mix of these two gas reservoirs \citep{2021ApJ...922L...6F,2022ApJ...928..144L}. Our intuition is here guided by idealised simulations of a wind interacting with a cloud showing that magnetic draping of the ICM over a projectile or over a galaxy \citep{2008ApJ...677..993D,2010NatPh...6..520P} and stripped gas from the cloud both contribute to the magnetic field in the tail \citep{2020MNRAS.499.4261S}.

Based on the maps in Fig.~\ref{Analyse10_Multipanel_vandB_testsim2HR_RotMatrix30deg_080_Movie}, we see that while ram-pressure stripping and cooling processes downstream the galaxy cause the magnetic field to be aligned with the tail, we see a $B_y$-component dominating over $B_x$ and $B_z$ in the immediate tail especially for the panels at $t\geq 2.70$ Gyr. Below we further quantify this result.

\begin{figure}
\centering
\includegraphics[width=0.79\linewidth]{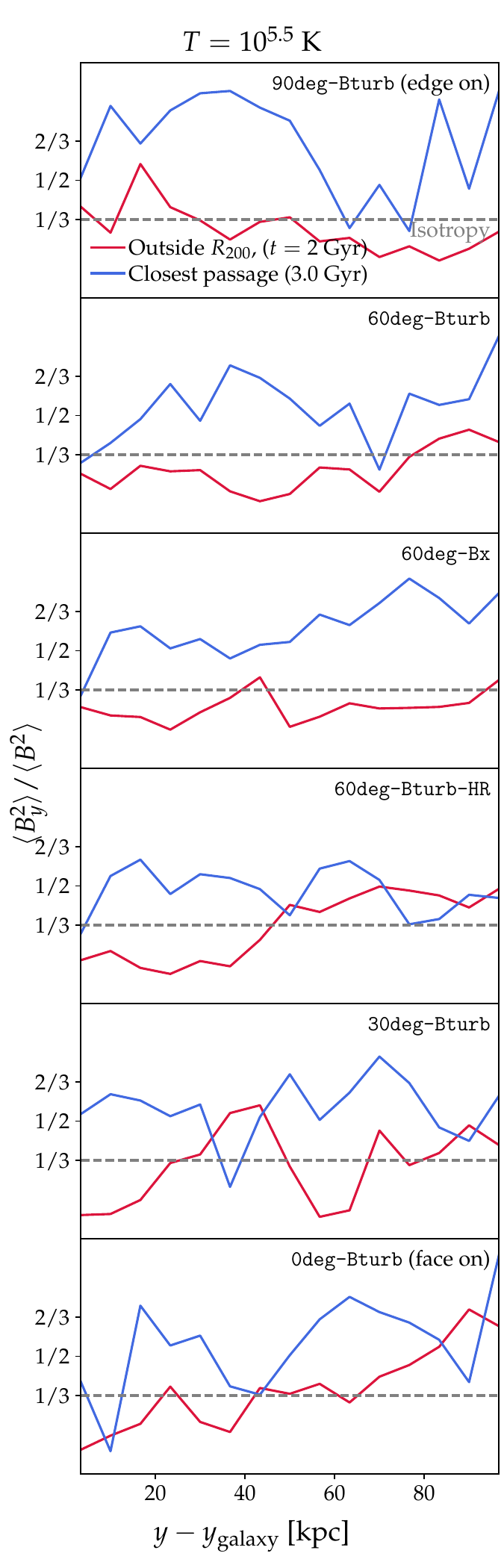}
\caption{For gas with a temperature of $10^{5.5\pm 0.25}$ K we measure whether the magnetic field is aligned with the jellyfish galaxy's tail (which is oriented along the $y$-direction). We call the field aligned if $\langle B_y^2\rangle/\langle B^2 \rangle > 1/3$. At 2.0 Gyr, which is prior to entering $R_{200}$, we see that $\langle B_y^2\rangle/\langle B^2 \rangle$ is close to the isotropic value ($1/3$). At the time of the closest passage to the cluster centre we frequently see an enhanced value of $\langle B_y^2\rangle/\langle B^2 \rangle$, which implies that the magnetic field aligns with the tail.}
\label{Analyse11_Multipanel_Bdir_SimplifiedVersion_logT}
\end{figure}

\begin{figure*}
\centering
\begin{minipage}{.48\textwidth}
\centering
\includegraphics[width=\linewidth]{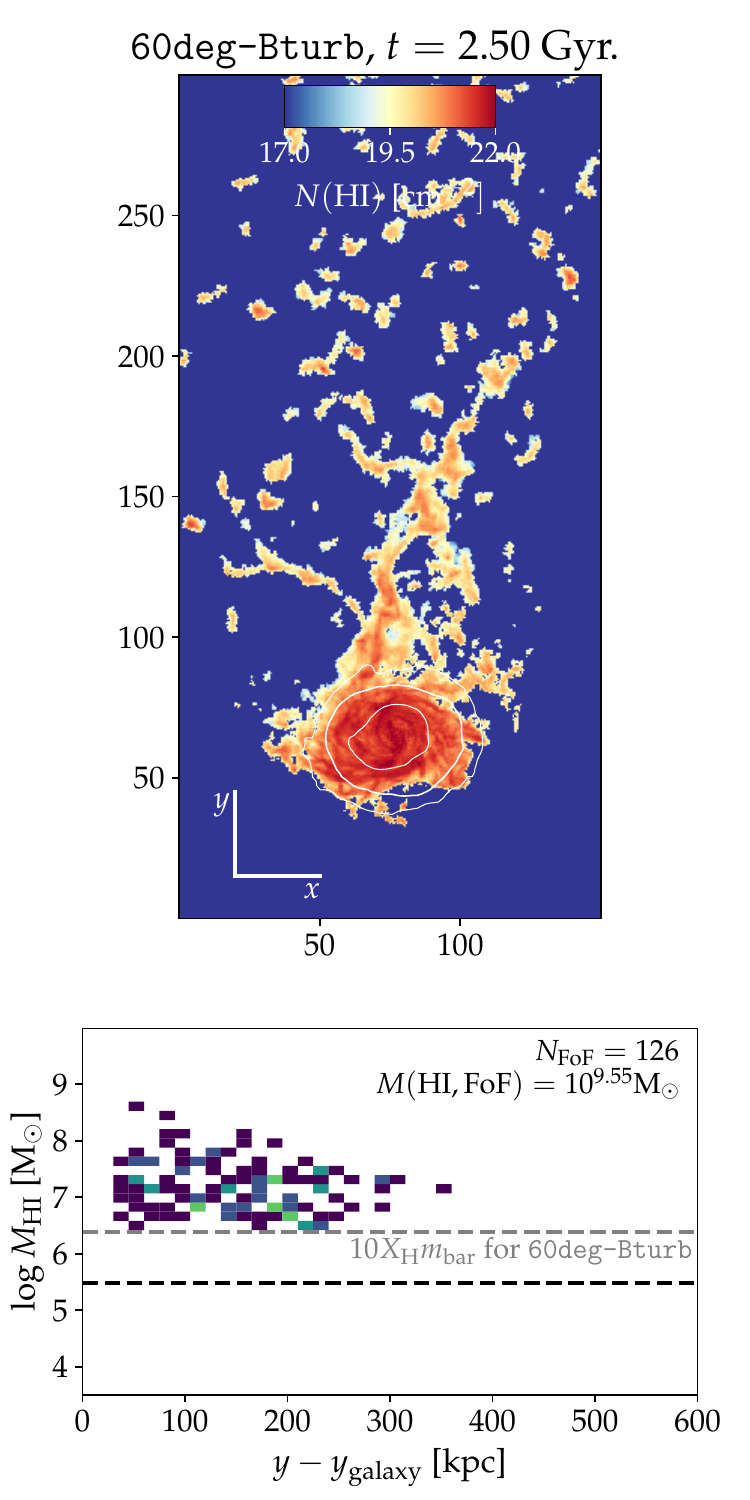}
\end{minipage}%
\begin{minipage}{.48\textwidth}
\centering
\includegraphics[width=\linewidth]{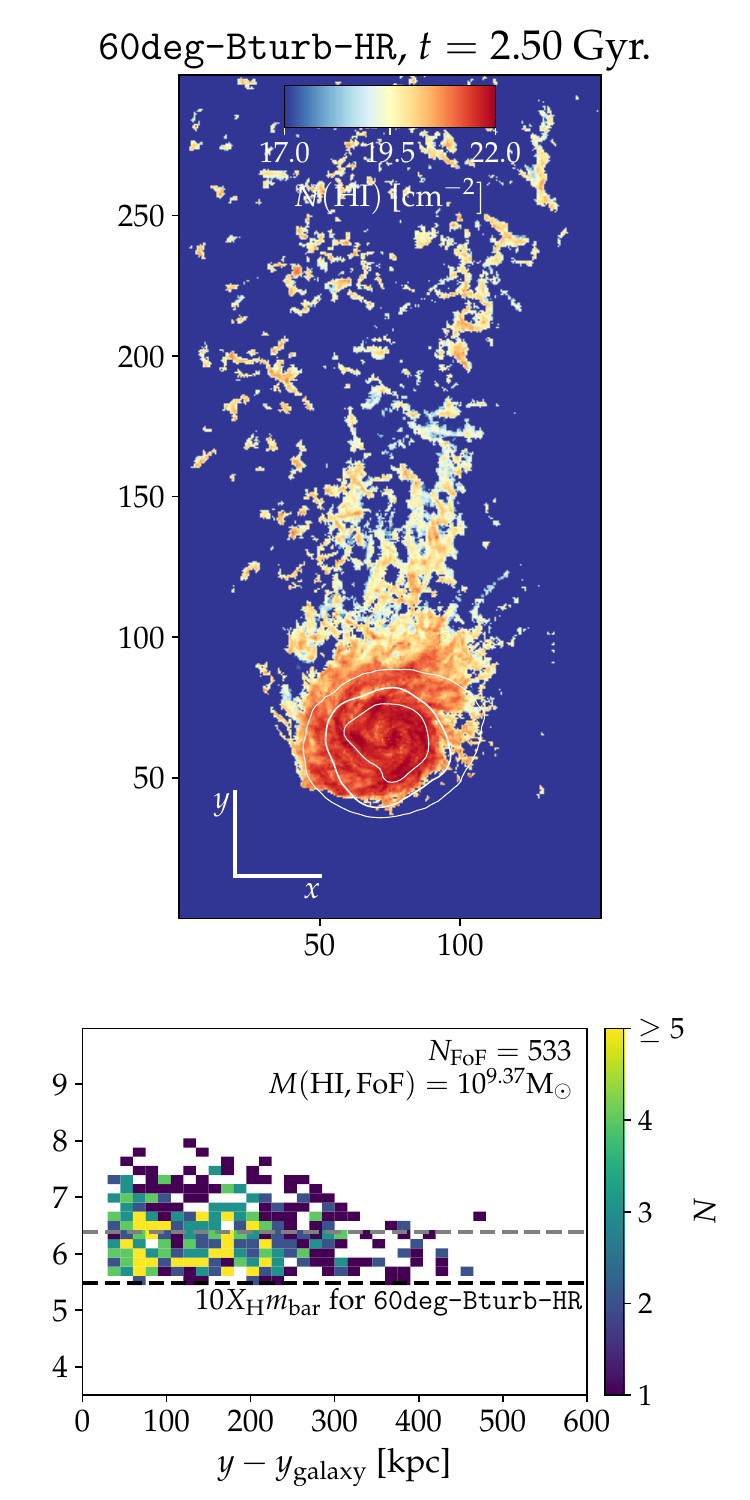}
\end{minipage}%
\caption{\emph{Upper panels}: At $t=2.50$ Gyr, which is before the central passage, we show \ion{H}{i} projection maps of the normal and high-resolution simulation with an inclination of 60 degrees (left and right, respectively). \emph{Lower panels}: we identify \ion{H}{i} clumps using a FoF algorithm. We plot the FoF mass as a function of the $y$-coordinate (the downstream distance from the galaxy centre). The dashed lines show the smallest possible \iona{H}{i} FoF mass, $10\,X_\rmn{H}m_\text{bar}$, i.e., each FOF group contains at least 10 gas cells. The high-resolution simulation contains more and less massive clumps showing that the clump size is limited by the resolution. It suggests that an unresolved thermal instability, \emph{shattering}, occurs in the tail.
}
\label{FigFoFAnalysisA}
\end{figure*}

\subsection{The alignment of the tail's magnetic field}\label{Sec:MagnetigFieldAlignment}

We now examine whether the magnetic field is aligned with the tail, by studying the ratio, $B_y^2/B^2$, which is 1/3 for an isotropic magnetic field distribution. 
If $B_y^2> B^2/3$, we regard the magnetic field as being aligned with the tail (which is oriented along the $y$-direction).

We compute the volume averaged values, $ \langle B_y^2 \rangle$ and $\langle B^2\rangle$, in linearly distributed bins in the $y$-direction. We use a weighting function, $G(\log T/\text{K})$, selecting gas with $10^{5.5\pm 0.25}$ K. This temperature range is selected to be close to the geometric mean of the wind and ISM temperature (inspired by the mixing temperature from \citealt{2018MNRAS.480L.111G}). We note that in Sect.~\ref{subsec:gasmaps}, we aimed at selecting gas at temperatures characteristic of \iona{H}{i} gas. Hence, with this different temperature selection we are studying slightly different magnetic field morphologies that are anchored in a different gas reservoir. We show in Appendix~\ref{SecMagneticFieldOrientation} that this temperature selection only affects details of the magnetic morphology indicating that the magnetic field is somewhat smoother for higher temperatures and shows a smaller standard deviation. Hence, this renders our main findings robust and independent on the specific temperature selection.

We plot $\langle B_y^2\rangle/\langle B^2 \rangle$ in Fig.~\ref{Analyse11_Multipanel_Bdir_SimplifiedVersion_logT} for all our simulations. We show an early time, $2.0$ Gyr, when the galaxy is outside $R_{200}$, and the time of the closest passage, $3.0$ Gyr. We see a complicated behaviour of $\langle B_y^2\rangle/\langle B^2 \rangle$ in all panels. When the galaxy is outside $R_{200}$ at $t=2.0$ Gyr the lowest inclination simulations ({\tt 0deg-Bturb}) has a lower $\langle B_y^2\rangle/\langle B^2 \rangle$-value in the immediate tail of the galaxy ($y-y_\text{galaxy}\lesssim 30$ kpc) in comparison to the highest inclination simulation ({\tt 90deg-Bturb}). This is because the initial magnetic of the disc in the former case has $B_y=0$ and in the latter case there is a non-zero component following Eq.~\eqref{Beq2}. Further downstream the interaction between the wind and the galaxy is increasingly complicated, and it is not possible to do a simple interpretation of the $\langle B_y^2\rangle/\langle B^2 \rangle$ ratios. We therefore simply address our main objective, which is to find out whether the interaction between ICM wind and the galaxy typically causes an aligned magnetic tail.

In simulation {\tt 90deg-Bturb}, where the wind hits the galaxy in an edge-on projection, $\langle B_y^2\rangle/\langle B^2 \rangle$ fluctuates around $1/3$ before the galaxy enters $R_{200}$. At the closest passage $\langle B_y^2\rangle/\langle B^2 \rangle$ is significantly boosted in the tail, implying that the magnetic field is aligned with the tail.

For all our simulations, we see the same trend; at the closest passage a large fraction of the tail has a $\langle B_y^2\rangle/\langle B^2 \rangle$ value being (i) larger than the isotropic value, and (ii) larger in comparison to when the galaxy was outside the cluster. This agrees well with the observations of the aligned magnetic tail observed in \citet{2021NatAs...5..159M} and is also seen in our simulations of a cold cloud being crushed by a hot wind \citep{2020MNRAS.499.4261S}.

The simulation where these trends are clearest is the one with an initially uniform magnetic field oriented perpendicular to the wind velocity ({\tt 60deg-Bx}). The uniform magnetic field in the wind makes the tail particularly ordered and aligned. 

In Appendix~\ref{SecMagneticFieldOrientation} we carry out additional detailed analysis, and show that this result is qualitatively unchanged for different temperature weighting functions. We specifically show plots with choices of $G$ selecting $10^{4.5\pm 0.25}$, $10^{5.5\pm 0.25}$ and $10^{6.5\pm 0.25}$ K gas. We also show more snapshot times to demonstrate how the ratio behaves during the infall into the cluster. In most cases the magnetic field is well aligned with the tail, and the only notable exception is for the edge-on galaxy, where we only see an aligned magnetic field at or after the central passage.

\subsection{Shattering in the tail}\label{subsec:shattering}

We now proceed to study cooling-induced shattering, which is expected to make the dense gas in the tail fragment either to a physical scale given by the cooling length \citep{2018MNRAS.473.5407M,2020MNRAS.499.4261S} or to the spatial resolution limit of the simulation (if the cooling length is not well-resolved).

In Fig.~\ref{FigFoFAnalysisA} we study the shattering in the tail at a time of 2.50 Gyr. This time is chosen because the tail is very well visible, which can also be seen in Fig.~\ref{Analyse10_Multipanel_vandB_testsim2HR_RotMatrix30deg_080_Movie}. We compare our normal resolution simulation ({\tt 60deg-Bturb}) to the corresponding high resolution version ({\tt 60deg-Bturb-HR}). The \ion{H}{i} column density projection reveals a rich structure of clumps in the tail of both simulations, but there seems to be more and smaller clumps in the high-resolution simulation.

To quantify the fragmentation we identify friends-of-friend (FoF) groups in the tail. We use a simple python implementation of a FoF finder algorithm, but for more information about FoF algorithms we refer to \citet{1985ApJ...292..371D} and \citet{2001MNRAS.328..726S}. We identify FoF groups with a density of $n(\ion{H}{i})\geq n_\text{FoF} =0.01 $ cm$^{-3}$. We hence set the FoF linking length to $l_\text{FoF}\equiv (2 m_\text{bar} X_\text{H}/[m_\text{p} n_\text{FoF}])^{1/3}$, where $X_\text{H}=0.76$ is the universal Hydrogen mass fraction, $2 m_\text{bar}$ is the largest mass allowed by our refinement criterion, and $m_\text{p}$ is the proton rest mass. Since our main aim is to characterise the tail, we only allow cells with being at least 40 kpc downstream from the galaxy ($y>y_\text{galaxy}+40$ kpc) to be included in a FoF group. We furthermore require a FoF group to contain at least 10 cells.

In both simulations we see \ion{H}{i} clumps hundreds of kpc downstream from the galaxy. At high resolution we notice the existence of clumps with lower masses in comparison to the normal resolution simulation. In the lower panels we overplot a dashed line showing the smallest possible \ion{H}{i} mass ($10X_\rmn{H} m_\text{bar}$) of a FoF group consisting of 10 gas cell according to our refinement criterion. We see that the lowest mass FoF group is just above this threshold for each simulation, so the gas clumps are fragmenting to the grid-scale in both simulations. 

In case of perfect numerical convergence, we would identify the same number of FoF groups in the two simulations (barring Poisson noise) and the same amount of gas mass locked up in cold clumps. We do, however, see many more groups in our high-resolution simulation in comparison to the normal resolution simulation (533 versus 126). This confirms the visual impression that there are more small-scale clumps in the high resolution simulation. This could be a consequence of stripped ISM gas having more resolved structure already before the gas moves into the tail. Alternatively, it could be a consequence of shattering and thermal instability in the tail.

In a follow-up paper (Sparre, Pfrommer, Puchwein in prep.) we trace gas flows in our simulations, and find that the majority ($>60$ per cent in {\tt 60deg-Bturb-HR}) of FoF groups downstream from the galaxy at this snapshot has never been cold within the ISM during the simulation, implying that ISM stripping is not the dominant channel for FoF group formation. Instead shattering and thermal instability is expected to play a larger role.

We quantify the \iona{H}{i} mass in the tail by summing up the total amount of \ion{H}{i} in our FoF groups. The result is listed in Fig.~\ref{FigFoFAnalysisA}. At the time of our selected snapshot the tail of the high-resolution simulation has a 0.18 dex lower mass in comparison to our normal-resolution simulation. This is consistent with more gas remaining in the disc in the high-resolution simulation during the galaxy-ICM interaction, which we saw in Fig.~\ref{Analyse8_MassEvolutionEvolution}. This characteristic is also seen at $t=2.0$, $2.25$ and $2.5$ Gyr, which are during the infall into the cluster, where our high-resolution simulation has a \iona{H}{i} tail with a mass being lower by 0.14, 0.10 and 0.18, in comparison to the normal-resolution simulation, respectively. This difference in tail mass is thus not a temporary feature, but instead a characteristic of the resolution dependence of the simulations. At $t=3.0$ -- the time of the nearest cluster passage -- this trend reverses and we see a 0.10 dex higher \iona{H}{i} mass in the high-resolution simulation.

\section{Discussion -- Jellyfish tails as physics laboratories}\label{Sec:Discussion}

\subsection{A feedback loop in the tail?}

\citet{2021NatAs...5..159M} found observational evidence for an aligned magnetic tail in a jellyfish galaxy, and speculated that a feedback loop could exist between (i) an aligned magnetic field driving gas into star-forming clouds, and (ii) star-forming clouds increasing magnetic turbulence around them and hence slowing down gas accretion. This feedback-cycle was named the \emph{draping scenario}.  

Our finding that the magnetic field orientation aligns with the jellyfish tail for a range of different simulations with different inclinations supports the overall initial assumptions of this scenario. It furthermore suggests that many more jellyfish galaxies should have a magnetic field aligned with the tail.

What we cannot address in our simulations is to what extent star-formation in the tail is supported by the aligned magnetic field or, alternatively, suppressed by magnetic turbulence injected by recently formed stars. This is because our ISM and star-formation model \citep{2003MNRAS.339..289S} uses a subgrid prescription for the interaction between cold star-forming clouds and their hot volume-filling surroundings, instead of explicitly resolving them. A consequence is that the impact of large-scale magnetic fields on star-forming regions is not well captured by the model. A next step in addressing the draping scenario would be to run our simulation's initial conditions with a resolved ISM model \citep[e.g.,][]{2019MNRAS.489.4233M,2021MNRAS.506.3882S,2023MNRAS.519.3154H}.

\subsection{Multi-wavelength mock observations}

In addition to completing new simulations with an explicit ISM model, there are several other science questions, which could be addressed in future work. For example, producing realistic multi-frequency observables such as H$\upalpha$ emission, X-ray emission \citep[using tools from e.g.,][]{2012MNRAS.420.3545B,2013MNRAS.428.1395B} and radio synchrotron emission \citep[as detected for jellyfish galaxies by e.g.][]{2023arXiv231020417R,2023A&A...675A.118I} produced by realistic cosmic ray electron spectra \citep{2019MNRAS.488.2235W,2020MNRAS.499.2785W,2021MNRAS.505.3273W,2021MNRAS.508.4072W,2021ApJS..253...18O,2023MNRAS.519..548B}. Realistic mock observations would, for example, demonstrate whether and how the shattering processes in the tail can be observed.

Simulations with a physical treatment of molecular gas would also be useful for unveiling the nature of multiphase gas below $10^4$ K. Molecular processes have been explored in cloud-crushing simulations \citep{2021MNRAS.505.1083G,2022MNRAS.510..551F}, but less in jellyfish galaxy simulations.

\section{Conclusion}\label{Sec:Conclusion}

We have presented windtunnel simulations of a jellyfish galaxy interacting with the ICM of a galaxy cluster. We have used a time-dependent, turbulent magnetic ICM wind, which enables us to quantify how the magnetic field in the tail behaves. Our main conclusions are the following:
\begin{itemize}
\item The interaction between the ICM wind and the galaxy causes a starburst at the time of the nearest cluster passage. A galaxy that is hit edge-on by the wind has a stronger starburst in comparison to a galaxy in a face-on orientation. The reason is that gas is funnelled towards the centre and compressed for the edge-on galaxy. We varied the disc orientation and found the SFR peak is gradually increasing for a $0^\circ$ (face on), $30^\circ$,$60^\circ$ and $90^\circ$ (edge on) inclination of the galaxy.
\item The gas stripping rate, especially at the time of the nearest passage, is gradually decreasing for the simulations with $0^\circ$ (face on), $30^\circ$, $60^\circ$ and $90^\circ$ (edge on), because of the higher cross section for a face-on galaxy in comparison to an edge-on galaxy.
\item The \iona{H}{i} gas in the tail of the galaxy exhibits a fountain flow, where the projected $\varv_y$ velocity component ($y$ is the tail direction) traces the rotation pattern of the galaxy itself. A fraction of gas in the tail is re-accreted back to the galaxy. In all our simulations, except for the case where the wind hits the galaxy face-on, such fountain flows are existing in the initial stages of the galaxy-ICM interaction after the galaxy has entered $R_{200}$, and before the nearest cluster passage. In the face-on case a fountain flow does not arise, because the rotation pattern of the galaxy has zero contribution in the direction of the tail.

\item At the time of the nearest cluster passage ram pressure stripping is so strong, so that $\varv_y$ is dominated by stripping and the fountain flow ceases to exist.

\item Independent of whether fountain flows or gas stripping signatures are present in the tail our statistical analysis reveals a magnetic field aligned with the tail. This is consistent with radio observations of the jellyfish galaxy JO 206 \citep{2021NatAs...5..159M} and in line with earlier simulation results of a cold cloud interacting with a wind \citep{2020MNRAS.499.4261S}.

\item By comparing simulations with mass resolutions different by a factor of eight, we find that the gas clumps in the jellyfish tail are not resolved at our normal resolution level. Using an FoF analysis, we confirm the visual impression of the high-resolution simulation exhibiting a richer structure in comparison to the normal-resolution simulation. This can be the result of shattering and/or ram-pressure stripping of a structured ISM.

\end{itemize}

This realistic simulation setup of a jellyfish galaxy has an enormous potential for serving as a laboratory to understand the star formation process in these extremely hostile environments of jellyfish tails that are exposed to a fast and hot wind. In the future, we plan on extending this setup by moving towards a multiphase ISM and by modelling the non-thermal cosmic ray electron and proton populations.

\section*{Acknowledgements}

We thank the referee for insightful comments. We thank Ivan Markin, Nikos Sagias and Timon Thomas for useful discussion. MS and CP acknowledge support by the European Research Council under ERC-AdG grant PICOGAL-101019746.

\section*{Data availability}
The data and scripts for this article will be shared on reasonable request to the corresponding author. The \textsc{arepo} code is publicly available.

\footnotesize{
\bibliographystyle{mnras}
\bibliography{ref1}
}

\appendix
%

\section{Magnetic alignment in different temperature plasma}\label{SecMagneticFieldOrientation}

In Sect.~\ref{Sec:MagnetigFieldAlignment} and Fig.~\ref{Analyse11_Multipanel_Bdir_SimplifiedVersion_logT} we studied the magnetic field orientation through the quantity, $\langle B_y^2\rangle/\langle B^2 \rangle$. For $10^{5.5\pm 0.25}$ K gas we found the trend that the magnetic field is preferentially aligned with the tail of the galaxy, i.e. $\langle B_y^2\rangle/\langle B^2 \rangle$>1/3, at the closest passage of the galaxy through the cluster. We here extend the analysis by adding analysis of snapshots at times 2.25, 2.5 and 2.75 Gyr -- at these times the galaxy has entered $R_{200}$ of the cluster, but the closest passage has not yet been reached (see Fig.~\ref{Fig10_PlotClusterWind}). We furthermore vary the weighting function, G used to calculate $\langle B_y^2\rangle/\langle B^2 \rangle$, to include either $10^{4.5\pm 0.25}$, $10^{5.5\pm 0.25}$ and $10^{6.5\pm 0.25}$ K gas.

By comparing the 2.0 and 3.0 Gyr snapshots (before entering $R_{200}$ and the closest cluster passage, respectively) we see that $\langle B_y^2\rangle/\langle B^2 \rangle$ is frequently larger than $1/3$ independent of the temperature of the gas. Our result that the magnetic field is aligned with the tail is therefore not very sensitive to the gas temperature within our studied range.

Studying our high-resolution simulation, {\tt 60deg-Bturb-HR}, we see that  $\langle B_y^2\rangle/\langle B^2 \rangle$ is mostly larger than $1/3$ for all the studied times. Especially in the immediate tail at $y_\text{galaxy, IC}=20-40$ kpc, we see that the interaction causes alignment above what is seen at the earliest shown snapshot. The same is true for $10^{6.5\pm 0.25}$ K gas in our simulation {\tt 30deg-Bturb}, but for lower temperatures this result is less clear. 

For the face-on inclination simulation, {\tt 0deg-Bturb}, we see an aligned field in most of the locations downstream at all temperatures and times. For the simulation with an edge-on inclination, {\tt 90deg-Bturb}, an aligned magnetic field is seen close to the central passage (at times of 2.75 and 3.0 Gyr), but less so earlier in the galaxy's orbit. Hence, magnetic fields are not in all circumstances aligned with the tail, but based on the analysis in this section, it is likely the case.

\begin{figure*}
\centering
\includegraphics[width=\linewidth]{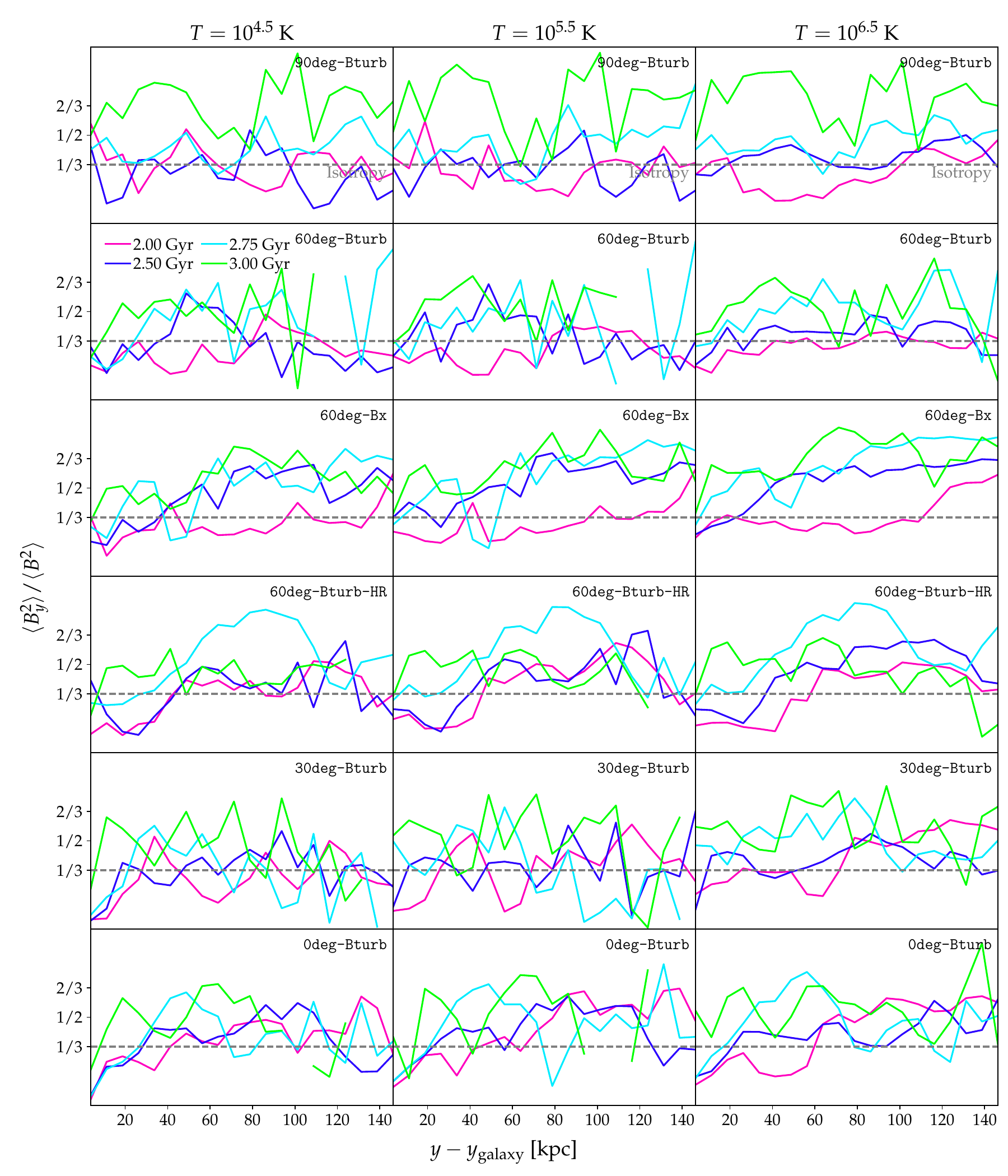}
\caption{The orientation of the magnetic field, which we also studied in Fig.~\ref{Analyse11_Multipanel_Bdir_SimplifiedVersion_logT}. Here we show four different times and probe a wider range of temperatures; $10^{4.5\pm 0.25}$, $10^{5.5\pm 0.25}$ and $10^{6.5\pm 0.25}$ K.}
\label{Analyse11_Multipanel_Bdir_logT}
\end{figure*}

\bsp	
\label{lastpage}
\end{document}